\documentstyle[preprint,epsfig,eqsecnum,aps]{revtex}

\begin{document}

\title{ {USM-TH-120}\\
The  $(\mu^-,\mu^+)$ conversion in nuclei as a probe of new physics}

\author{
Fedor \v Simkovic\footnote{On leave of absence from Department of
Nuclear Physics,  Comenius University, Mlynsk\'a dolina F1,
SK--842 15 Bratislava, Slovakia}, Amand Faessler}

\address{
Institut f\"ur Theoretische Physik, Universit\"at
T\"ubingen, Auf der Morgenstelle 14, D-72076 T\"ubingen, Germany}

\author{
Sergey Kovalenko\footnote{On leave of absence from Join Institut
for Nuclear Research, Dubna, Russia}, Ivan Schmidt}

\address{ Departamento de F\'\i sica, Universidad
T\'ecnica Federico Santa Mar\'\i a, Casilla 110-V, Valpara\'\i so, Chile}

\date{\today}
\maketitle
\draft
\begin{abstract}
A detailed study of the  muonic analogue of neutrinoless  double
beta decay, $(\mu^-,\mu^+)$ conversion, has been carried out for
the A=44 nuclear system. We studied several lepton number
violating (LNV) mechanisms potentially triggering this process:
exchange by light and heavy Majorana neutrinos as well as exchange
by supersymmetric particles participating in R-parity violating
interactions. The nuclear structure has been taken into account
within the renormalized Quasiparticle Random Phase Approximation
method. To our knowledge, this is the first realistic  treatment
of nuclear structure aspects of the $(\mu^-,\mu^+)$ conversion. We
estimated the rate of this process utilizing the existing
experimental constraints on the parameters of the underlying LNV
interactions and conclude that the $(\mu^-,\mu^+)$ conversion is
hardly detectable in the near future experiments.
\end{abstract}

\section{Introduction}

Muonic atoms are known to be a suitable laboratory for studying
lepton flavor and lepton number violation (LFV, LNV)
\cite{marc,miss,mupo,ko94,ko97,mue,kos01,sim97,sind}. The
main attention has been previously paid to the reactions of
muon--electron [$(\mu^-,e^-)$]  and muon--positron [$(\mu^-,e^+)$]
conversions in nuclei. The new forthcoming experiments with
stopped muon beams  MECO \cite{meco} at BNL and PRISM at KEK
\cite{prime} are going to substantially improve sensitivity to
different LFV/LNV decay channels. The aimed intensity of muon beam
from the proton machines is about four orders of magnitude higher
than that available at present. This substantial experimental
progress is raising the question on observability of other muonic
LFV/LNV processes as yet not addressed
experimentally.

In the present paper we are studying the $(\mu^-,\mu^+)$ conversion
in nuclei
\begin{equation}
\mu^-_b ~ + ~ (A,Z) ~\rightarrow ~ (A,Z-2)~ + \mu^+,
\label{eq.1}
\end{equation}
the muonic analogue of neutrinoless double decay, first proposed
in Ref.  \cite{miss}. The $\mu_b$ is a bounded muon in a 1S atomic
state. This process has never been searched for experimentally.
The rough estimates of Ref. \cite{miss} within some extensions of
the standard model (SM) show that its branching ratio $R_{\mu^+
\mu^-} = {\Gamma_{\mu^- \mu^+}}/{\Gamma_{\mu \nu_\mu}}$ to
ordinary muon capture does not exceed $10^{-18}$. Although this is
a very small number it does not look unreachable for future
experiments. In fact, the expected sensitivity of the forthcoming
experiments MECO and PRIME to another neutrinoless nuclear
process, $(\mu^-,e^-)$ conversion, is of the same order of
magnitude $\sim 10^{-18}$. Despite significant differences in
searching strategies for the $(\mu^-,e^-)$  and the
$(\mu^-,\mu^+)$ conversion processes as well as in the
technological treatment of the corresponding nuclear targets, the
above observation might be encouraging for future searches of
nuclear $(\mu^-,\mu^+)$ conversion. The results of our present
study lead to a less optimistic conclusion. We argue that the rate
of this process is likely by at least 4 orders of magnitude less
than that given in  Ref. \cite{miss}, and thus its possible
observation is shifting to a quite distant future.

We study the $(\mu^-,\mu^+)$ conversion in nuclei within
the most conventional LNV extensions of the SM with light and heavy
Majorana neutrinos as well as in the minimal supersymmetric SM extension
with R-parity violation. We carried out a detailed analysis of the nuclear
structure effects in the $(\mu^-,\mu^+)$ conversion
within the renormalized quasiparticle random phase
approximation (RQRPA) \cite{schw,fae,woda}.
The theoretical upper bounds
on the branching ratios are evaluated
using the current constraints on the relevant LNV parameters.

The paper is organized as follows. In Section 2 we discuss
kinematical conditions and admissible nuclei for searching for the
$(\mu^-,\mu^+)$ conversion. There we also give the $(\mu^-,\mu^+)$
conversion rate formulas and specify their parameters. Sections
3,4,5 deal with specific mechanisms of $(\mu^-,\mu^+)$ conversion.
We determine the corresponding LNV parameters in terms of the
fundamental parameters of the considered models, derive the
$(\mu^-,\mu^+)$ transition operators and their nuclear matrix
elements. In Section 6 we calculate these nuclear matrix elements
for the long-living radioactive isotope $^{44}Ti$ in the RQRPA.
Then we derive upper bounds on the  $(\mu^-,\mu^+)$ conversion in
$^{44}Ti$ using existing experimental constraints on the
parameters of the studied LNV models and discuss prospects for its
observability in future experiments. The summary is given in
Section IV.

\section{Theoretical preliminaries}

The $(\mu^-,\mu^+)$ nuclear conversion violates total lepton
number conservation by two units and, therefore, is forbidden in
the standard model. Thus, its experimental observation, as well as
the observation of any other LNV process, would be an unambiguous
signal of physics beyond the SM. In general, a lepton number
violating process may proceed if and only if neutrinos are
Majorana particles with non-zero masses \cite{theorem}
 which can not be accommodated into the SM.

The underlying LNV mechanisms for different LNV processes
are very similar and can be adopted from the previously studied
case of neutrinoless double beta decay ($0\nu\beta\beta$-decay)
\cite{dbd-new}.
However, kinematically different LNV processes are rather different.

As observed in Ref. \cite{miss}, the energy conservation requirement
for the  $(\mu^-,\mu^+)$ conversion,
\begin{equation}
T = E_i - E_f - \varepsilon_b > 0
\label{eq.2}
\end{equation}
is not fulfilled by any stable nucleus (A,Z). Here, T is the
kinetic energy of the outgoing muon, $E_i$ ($E_f$) is the ground
state energy of initial (final) nucleus and $\varepsilon_b$ is the
muon binding energy in the muonic atom. Thus, experimental
searches for the $(\mu^-,\mu^+)$ conversion are possible only with
radioactive isotopes. The only feasible target nucleus would be
the isotope $^{44}Ti$ with sufficiently long lifetime of 47 years
\cite{miss}.

Following Ref.  \cite{miss} we write down the rate of the $(\mu^-,\mu^+)$
conversion as
\begin{eqnarray}
\Gamma_{\mu\mu}^{(i)} = \frac{1}{\pi}~ E_{\mu^+}~ p_{\mu^+}~
F(Z-2,E_{\mu^+})~
c_{\mu\mu} ~
|{\cal M}^{\Phi}_{(i)}|^2 ~
\left|\eta_{(i)}\right|^2,
\label{eq.3}
\end{eqnarray}
where $E_{\mu^+} = m_\mu - \varepsilon_b + E_i - E_f$, $p_{\mu^+}
= \sqrt{E^2_{\mu^+} - m_\mu^2 }$ ($m_\mu$ is the mass of muon),
$c_{\mu\mu}~ = 2 G_F^4 \left({(m_e m_\mu)}/{(4 \pi m_\mu
R)}\right)^2 g_A^4$. For the isotope $^{44}Ti$ one has:
$\varepsilon_b = 1.28~MeV$, $p_{\mu^+} = 18.5~MeV/c$. The index
$(i)$  denotes the specific mechanism of the $(\mu^-,\mu^+)$
conversion that we will study in the subsequent sections. In Eq.
(\ref{eq.3}) the factors $\eta_{(i)}$ and ${\cal M}^{\Phi}_{(i)}$
are the LNV parameters and the nuclear matrix elements,
respectively, associated with these mechanisms. The latter involve
smearing over the muon wave function $\Phi(r)$.
Finally, the relativistic Coulomb factor 
$F(Z-2,E_{\mu^+})$ is \cite{DOI85}
\begin{equation}
F(Z-2,E_{\mu^+}) = \left(\frac{\Gamma (5)}{ \Gamma(2\gamma +1)}\right)^2
(2 p_{\mu^+} R)^{2(\gamma -1)} |\Gamma (\gamma - i y)|^2 e^{-\pi y},
\end{equation}
where $\gamma = \sqrt{1 - (\alpha (Z-2))^2}$, 
$y =  \alpha (Z-2) E_{\mu^+}/p_{\mu^+}$.  R is the nuclear radius.

In the experiments with stopped muon beams one measures
the branching ratio $R_{\mu^+\mu^-}^{(i)}$ of the $(\mu^-,\mu^+)$
conversion
\begin{eqnarray}
R^{(i)} = \frac{\Gamma_{\mu\mu}^{(i)}}{\Gamma_\mu}
\label{eq.31}
\end{eqnarray}
with respect to the ordinary muon capture
$\mu^-_b + (A,Z) \rightarrow (A,Z-1) + \nu_\mu$.
The rate $\Gamma_{\mu}$ of this
SM allowed process can be calculated with the Primakoff formula \cite{pri}:
\begin{eqnarray}
\Gamma_\mu = \frac{1}{2\pi} m^2_\mu  (G_F \cos \theta_c)^2 <\Phi_\mu >^2 Z
[G^2_V + 3 G^2_A + G_P^2 - 2 G_A G_P] f(Z,A),
\label{eq.33}
\end{eqnarray}
The Pauli blocking factor for $^{44}Ti$ takes the
value $f(22,44) = 0.16$  \cite{leon}.
The numerical value of the quadratic combination of
the nucleon weak coupling constants is
$[G^2_V + 3 G^2_A + G_P^2 - 2 G_A G_P] \approx 5.9$.

In  the subsequent sections we derive the LNV parameters
$\eta_{(i)}$ and the nuclear matrix elements ${\cal
M}^{\Phi}_{(i)}$ for three examples of  mechanisms of
$(\mu^-,\mu^+)$ conversion based on the light and heavy Majorana
neutrino exchange as well as on the LNV couplings of R-parity
violating SUSY.

\section{Light Majorana neutrino exchange mechanism}

This mechanism is described by the diagrams in Fig. 1(a,b). In
this case the LNV parameter for Eq. (\ref{eq.3}) is given by
\begin{eqnarray}
\eta_{\nu} = \left|\frac{<m_\nu >_{\mu\mu}}{m_e}\right|^2
\end{eqnarray}
where the effective muon neutrino mass is
\begin{equation}
<m_\nu >_{\mu\mu} ~= ~ \sum^{light}_k~ (U_{\mu k})^2 ~ \xi_k ~ m_k.
\label{eq.4}
\end{equation}
Here, $U$ is the unitary neutrino
mixing matrix relating the weak and Majorana mass eigenstates of neutrinos.
The mass eigenvalues and the factor of relative CP-phase are  $m_k$
and  $\xi_k$ respectively.
The summation in Eq. (\ref{eq.4}) runs over the light neutrino
mass eigenstates with the masses $m_k$ much less than the typical
energy scale of  the $(\mu^-,\mu^+)$ conversion ($m_k \ll 100$ MeV).

We calculate the nuclear matrix element ${\cal M}^{\Phi}_{\nu}$
for this mechanism in the standard way, using the non-relativistic
impulse approximation \cite{DOI85}. The final result can be
written in terms of the Gamow-Teller $M^{\Phi}_{GT(\nu)}$ and the
Fermi $M^{\Phi}_{F(\nu)}$ nuclear matrix elements as
\begin{equation}
{\cal M}^{\Phi}_{\nu} = - \frac{M^{\Phi}_{F(\nu)}}{g^2_A}
+ M^{\Phi}_{GT(\nu)}
\label{eq.5}
\end{equation}
with
\begin{eqnarray}
&&M^{\Phi}_{F(\nu)} = \frac{4 \pi R}{(2 \pi)^3}
\int \frac{d\vec{q}}{2 q} \times \nonumber \\
&&\hspace*{6mm} \sum_n \left(
\frac{
\langle 0^+_i|\sum_l \tau^+_l
e^{-i \vec{q}\cdot{\vec{r}}_l} |n\rangle \langle n|
\sum_m \tau^+_m
e^{i \vec{q}\cdot{\vec{r}}_m} \Phi (r_m) |0^+_f\rangle }
{q - E_b + E_n - E_i + i \varepsilon_n}
 + \right. \nonumber \\
&& \left.
~~~\frac{
\langle 0^+_i|\sum_m \tau^+_m
e^{i \vec{q}\cdot{\vec{r}}_m} \Phi (r_m)
|n\rangle \langle n|\sum_l \tau^+_l
e^{-i \vec{q}\cdot{\vec{r}}_l}
|0^+_f\rangle }
{q + E_{\mu^+} + E_n - E_i + i \varepsilon_n}
  \right),
\label{eq.6}
\end{eqnarray}
\begin{eqnarray}
&&M^{\Phi}_{GT(\nu)} = \frac{4 \pi R}{(2 \pi)^3}
\int \frac{d\vec{q}}{2 q} \times \nonumber \\
&&\hspace*{6mm} \sum_n \left(
\frac{
\langle 0^+_i|\sum_l \tau^+_l {\vec{\sigma}}_l
e^{-i \vec{q}\cdot{\vec{r}}_l} |n\rangle \cdot \langle n|
\sum_m \tau^+_m {\vec{\sigma}}_m
e^{i \vec{q}\cdot{\vec{r}}_m} \Phi (r_m) |0^+_f\rangle }
{q - E_b + E_n - E_i + i \varepsilon_n}
 + \right. \nonumber \\
&& \left.
~~~\frac{
\langle 0^+_i|\sum_m \tau^+_m {\vec{\sigma}}_m
e^{i \vec{q}\cdot{\vec{r}}_m} \Phi (r_m)
|n\rangle \cdot \langle n|\sum_l \tau^+_l {\vec{\sigma}}_l
e^{-i \vec{q}\cdot{\vec{r}}_l}
|0^+_f\rangle }
{q + E_{\mu^+} + E_n - E_i + i \varepsilon_n}
 \right).
\label{eq.7}
\end{eqnarray}
Here,  $\Phi (r)$ is the radial part of
bound muon wave function in the 1S-state,
$E_b$ and $E_n$ are the energies of bound muon $\mu^-$ and
intermediate nuclear state, respectively.
The width of the n-th intermediate nuclear state,
denoted by $\varepsilon_n$, is  very small being determined only
by the electromagnetic and weak decay channels. For low lying nuclear
states it is much smaller than the corresponding intermediate state energy,
$\varepsilon_n \ll E_n$. This observation will allow us to
significantly simplify the calculations presented in Section 6.

Deriving the $(\mu^-,\mu^+)$ conversion matrix elements we assumed
that the outgoing muon $\mu^+$ is in $s_{1/2}$ state and,
therefore, $p_{\mu^+}R/3 \approx 0.1$. We also neglected the
contributions from the weak-magnetism and induced pseudoscalar
couplings of the nucleon current. These contributions correspond
to higher order terms of non-relativistic expansion and are not
important for our analysis. In fact, their inclusion may reduce
the values of the nuclear matrix elements only up to $20-30\%$
\cite{si99}.

The $(\mu^-,\mu^+)$ conversion matrix elements differ considerably
from the nuclear matrix elements of $0\nu\beta\beta$-decay
\cite{si99}. Note, first, that the value  of $(- E_b + E_n - E_i)$
entering the denominators of Eqs. (\ref{eq.6}), (\ref{eq.7}) is
negative, i.e., the nuclear matrix elements exhibit a singular
behavior. Therefore, the widths of the intermediate nuclear states
are of great importance, and the imaginary part of the
$(\mu^-,\mu^+)$ conversion matrix element can be large. We note
that this point was missed in Ref. \cite{miss}. A similar behavior
of the amplitude of the muon to positron conversion process has
been indicated only recently \cite{mupo}.

In order to simplify the numerical calculation of the
$(\mu^-,\mu^+)$ conversion matrix element we adopt two additional
approximations.\\
{\bf (i)} We assume that the 1S muon wave function varies very
little inside the nucleus $^{44}Ti$. Thus we apply the usual
approximation \cite{ko94}
\begin{equation}
|{\cal M}^{\Phi}_{\nu}|^2 ~=
<\Phi_\mu >^2 ~|{\cal M}_{\nu}|^2,
\label{eq.8}
\end{equation}
where the  muon average probability density over the nucleus is
\begin{equation}
<\Phi_\mu >^2 \equiv \frac{\int |\Phi_\mu (\vec{x})|^2 \rho(\vec{x}) d^3 x}
{\int \rho(\vec{x}) d^3 x}.
\label{eq.9}
\end{equation}
Here, $\rho(\vec{x})$ is the nuclear density. To a good
approximation it has been found that \cite{ko94}
\begin{equation}
<\Phi_\mu >^2 = \frac{ \alpha^3 m_\mu^3}{\pi} \frac{Z^4_{eff}}{Z},
\label{eq.10}
\end{equation}
i.e., the deviation from the behavior of the wave function at the
origin  has been taken into account by the effective proton number
$Z_{eff}$ ( $Z_{eff} = 17.5$ for $Z=22$ \cite{ko94}).\\
{\bf (ii)} We complete the sum over intermediate nuclear states
by closure after replacing $E_n$, $\varepsilon_n$
by some average values $\bar{E}$, $\bar{\varepsilon}$, respectively.

\begin{figure}[t]
\centerline{\epsfig{file=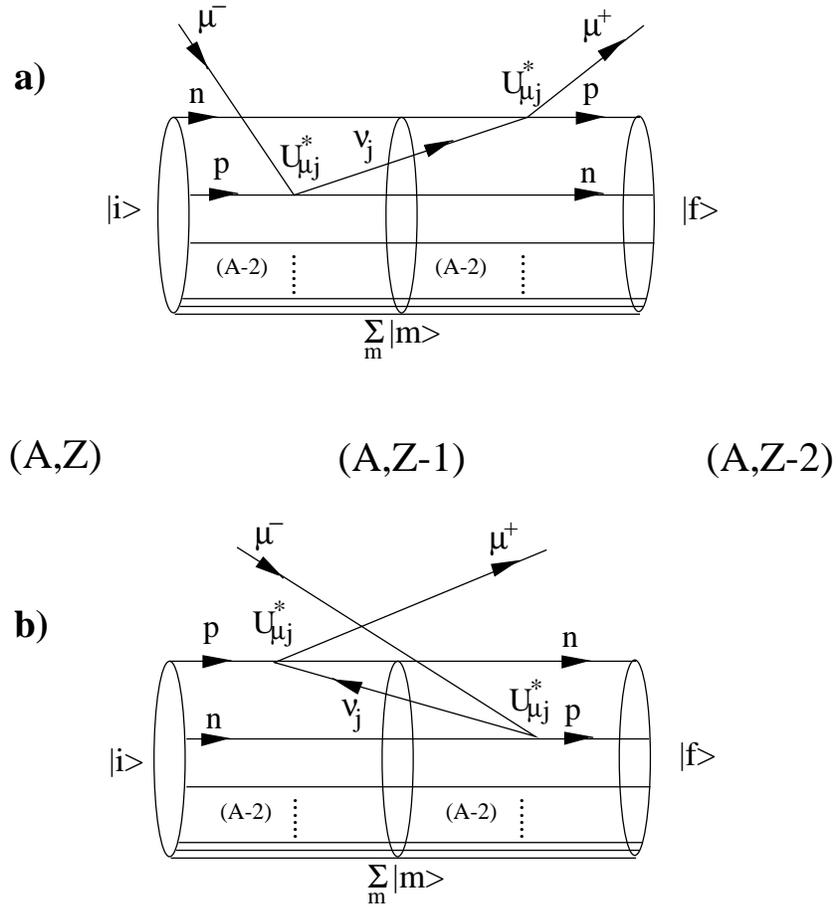,height=12.0cm}}
\vspace{2.cm}
\caption{The direct (a) and cross (b) Feynman diagrams of the
 $(\mu^-,\mu^+)$ conversion in nuclei mediated by Majorana neutrinos.
}
\label{fig.1}
\end{figure}

Then we obtain
\begin{equation}
{\cal M}_{\nu} = -\frac{M_{F(\nu)}}{g^2_A}
+ M_{GT(\nu)} = M^D + M^C,
\label{eq.11}
\end{equation}
with
\begin{eqnarray}
&&M_{F(\nu)} = M^D_{F} +  M^C_{F} \\
&&=
\langle 0^+_i|\sum_{kl} \tau^+_k \tau^+_l
\frac{R}{\pi} \int_0^\infty
\frac{j_0(q r_{kl})}{q - E_b + \bar{E} - E_i + i \bar{\varepsilon}}
f_V^2(q^2) q dq |0^+_f\rangle + \nonumber \\ \nonumber
&& ~\langle 0^+_i|\sum_{kl} \tau^+_k \tau^+_l
\frac{R}{\pi} \int_0^\infty
\frac{j_0(q r_{kl})}{q + E_{\mu^+} + \bar{E} - E_i + i \bar{\varepsilon}}
f_V^2(q^2) q dq |0^+_f\rangle ,
\label{eq.12}
\end{eqnarray}
\begin{eqnarray}
&&M_{GT(\nu)} = M^{D}_{GT} +  M^C_{GT} \\
&&=
\langle 0^+_i|\sum_{kl} \tau^+_k \tau^+_l \vec{\sigma_k}\cdot\vec{\sigma_l}
\frac{R}{\pi} \int_0^\infty
\frac{j_0(q r_{kl})}{q - E_b + \bar{E} - E_i + i \bar{\varepsilon}}
f_A^2(q^2) q dq |0^+_f\rangle  + \nonumber \\ \nonumber
&& ~\langle 0^+_i|\sum_{kl} \tau^+_k \tau^+_l
\vec{\sigma_k}\cdot\vec{\sigma_l}
\frac{R}{\pi} \int_0^\infty
\frac{j_0(q r_{kl})}{q + E_{\mu^+} + \bar{E} - E_i + i \bar{\varepsilon}}
f_A^2(q^2) q dq |0^+_f\rangle .
\label{eq.13}
\end{eqnarray}
Here we take the nucleon form factors in the dipole form $f_V(q^2)
= 1/(1+q^2/\Lambda^2_V)^2$ [$\Lambda^2_V = 0.71~(GeV)^2$],
$f_A(q^2) = 1/(1+q^2/\Lambda^2_A)^2$ [$\Lambda_A = 1.09~GeV$].
Note that the nuclear matrix elements $M_{F(\nu)}$ and
$M_{GT(\nu)}$ are normalized in the same way as the matrix
elements of the $0\nu\beta\beta$-decay process, allowing for their
direct comparison. We also separate the $(\mu^-,\mu^+)$ conversion
nuclear matrix elements into two terms associated with the direct
($M^{D}_{F}$, $M^{D}_{GT}$ and $M^D$) and the cross ($M^{C}_{F}$,
$M^{C}_{GT}$ and $M^C$) diagrams (see Fig. \ref{fig.1}). The
direct terms contain the denominators with singular behavior.

\section{Heavy Majorana neutrino exchange mechanism}

We assume that the neutrino mass spectrum include heavy Majorana
states $N$ with masses $M_k$ much larger than the typical energy
scale of the $(\mu^-,\mu^+)$ conversion, $M_k \gg 100$ MeV. These
heavy states can mediate this process according to the same
diagrams in Fig. 1(a.b) as the previous light neutrino exchange
mechanism. The difference is that the neutrino propagators in the
present case can be contracted to points and, therefore, the
corresponding effective transition operators are local unlike in
the light neutrino exchange mechanism with long range internucleon
interactions. As shown latter, the effect of nuclear structure in
both cases is very different.

The corresponding LNV parameter $\eta_{_N}$ is given by
\begin{equation}
\eta_{_N}
~= ~ \sum^{heavy}_k~ (U_{\mu k})^2 ~ {\xi}_k ~
\frac{m_p}{M_k}.
\label{eq.15}
\end{equation}
Here, $m_p$ is the mass of proton. We assume that the mass of
heavy neutrinos $M_k$ is large in comparison with their average
momenta ($M_k >> 1$ GeV) and that the muon wave function varyies
only slightly inside the nucleus.  Then the resulting nuclear
matrix element ${\cal M}_{_N}$ takes the same form as the
corresponding nuclear matrix element for the heavy neutrino mass
mechanism of $0\nu\beta\beta$-decay \cite{si99}. Separating the
Fermi (F), Gamow-Teler (GT) and the tensor (T) contributions we
write down
\begin{eqnarray}
{\cal M}_{_N} &=& - \frac{M_{F(N)}}{g^2_A}
+ M_{GT(N)} + M_{T(N)} \nonumber \\
&=&
\langle 0^+_i|\sum_{kl} \tau^+_k \tau^+_l
[ \frac{H_F(r_{kl})}{g^2_A} +
H_{GT}(r_{kl}) \sigma_{kl} - H_T(r_{kl}) S_{kl}]
|0^+_f\rangle ,
\label{eq.16}
\end{eqnarray}
where
\begin{equation}
S_{kl} = 3({\vec{ \sigma}}_k\cdot \hat{{\bf r}}_{kl})
       ({\vec{\sigma}}_l \cdot \hat{{\bf r}}_{kl})
      - \sigma_{kl}, ~~~ \sigma_{kl}=
{\vec{ \sigma}}_k\cdot {\vec{ \sigma}}_l.
\label{eq.17}
\end{equation}
${\bf r}_k$ and ${\bf r}_l$ are coordinates of nucleons undergoing
weak interaction.
${\bf r}_{kl} = {\bf r}_k-{\bf r}_l$,
$r_{kl} = |{\bf r}_{kl}|$,
$\hat{{\bf r}}_{kl} = {{\bf r}_{kl}}/{r_{kl}}$.
The radial part of the exchange potential is
\begin{equation}
H_I(r_{kl}) = \frac{2}{\pi} \frac{R}{m_p m_e}
\int_0^\infty j_0(q r_{kl}) h_I(q^2) q^2 dq ~~~~(I = F, ~GT, ~T),
\label{eq.18}
\end{equation}
with
\begin{eqnarray}
h_{F}(q^2) & = &   ~f^2_V({q}^{~2}) \nonumber \\
h_{GT}(q^2)  & = & f^2_A({q}^{~2}) ~[ ~1~ -~
\frac{2}{3}~ \frac{ {q}^{~2}}{ {q}^{~2} + m^2_\pi } ~+ ~
\frac{1}{3} ~( \frac{ {q}^{~2}}{ {q}^{~2} + m^2_\pi } )^2 ~]
~ +~ \frac{2}{3}~ \frac{f^2_M ({q}^{~2} ) {q}^{~2} }{4 m^2_p },
\nonumber \\
h_T (q^2) & = & f^2_A({q}^{~2})~ [~
\frac{2}{3}~ \frac{ {q}^{~2}}{ {q}^{~2} + m^2_\pi } -
\frac{1}{3}~ ( \frac{ {q}^{~2}}{ {q}^{~2} + m^2_\pi } )^2 ~]~
~+~ \frac{1}{3} ~\frac{f^2_M ({q}^{~2} ) {q}^{~2} }{4 m^2_p },
\label{eq.19}
\end{eqnarray}
Here, $m_\pi$ is the  mass of pion, $f_V(q^2)$, $f_A(q^2)$
are the nucleon form factors introduced in Eqs. (\ref{eq.12}),
(\ref{eq.13}) and $g_M({q}^{~2})= (\mu_p-\mu_n) f_V({q}^{~2})$.

\section{Trilinear R-parity breaking contribution}

In the SUSY models with R-parity non-conservation there are present
the LNV couplings which  may trigger the $(\mu^-,\mu^+)$ conversion.
Recall, that R-parity is a multiplicative
quantum number defined by $R=(-1)^{2S+3B+L}$ (S,B,L are spin, baryon
and lepton number). Ordinary particles have $R=+1$ while their
superpartners $R=-1$.
The LNV couplings emerge in this class of SUSY models from
the R-parity breaking part of the superpotential
\begin{equation}
W_{R_{p}\hspace{-0.8em}/\;:}=\lambda _{ijk}L_{i}L_{j}E_{k}^{c}+\lambda
_{ijk}^{\prime }L_{i}Q_{j}D_{k}^{c} + \mu_i L_i H_2,  \label{W-Rp}
\end{equation}
where $L$, $Q$ stand for lepton and quark $SU(2)_{L}$
doublet left-handed superfields, while $E^{c},D^{c}$ for
lepton and down quark singlet superfields.
Below we concentrate only on the trilinear $\lambda'$-couplings.
This is motivated by the fact that they give the dominant contribution
in the case of the $0\nu\beta\beta$-decay \cite{LH}.

\begin{figure}[t]
\centerline{\epsfig{file=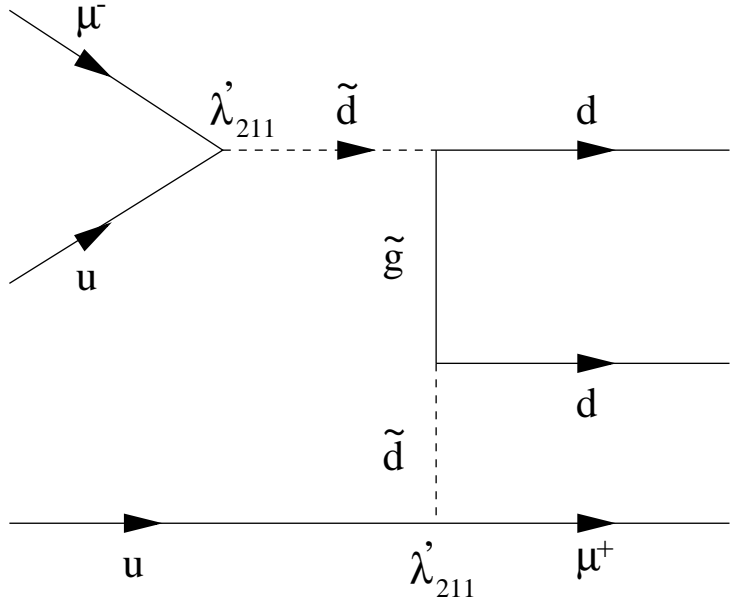,height=8.0cm}} \vspace{2.cm}
\caption{An example of the supersymmetric contribution to
$(\mu^-,\mu^+)$ conversion at the quark level} \label{fig.2}
\end{figure}

The $(\mu^-,\mu^+)$ conversion mechanisms in this model can be
derived in a close similarity with the $0\nu\beta\beta$-decay
\cite{FKSS97,FKS98a}. An example of the diagram representing such
a mechanism at the quark level is given in Fig. \ref{fig.2}. This
and all the other diagrams of this mechanism involve only the
${\lambda'}_{211}$ coupling.

\begin{figure}[t]
\centerline{\epsfig{file=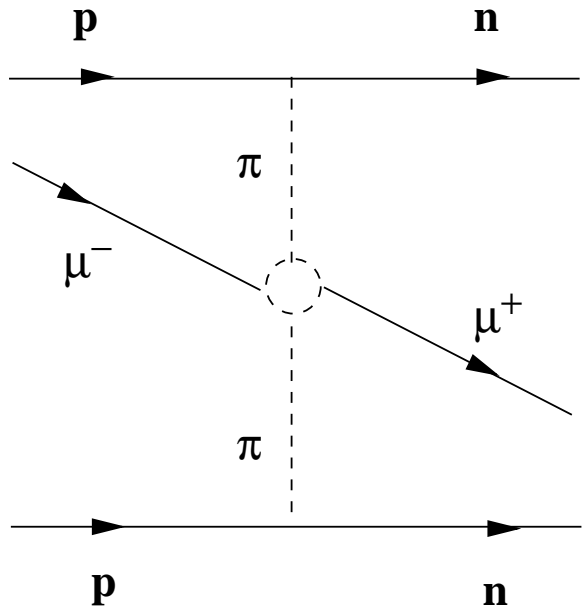,height=8.0cm}} \vspace{1.cm}
\caption{The pion-exchange mechanism of $(\mu^-,\mu^+)$ conversion
in nuclei. An example of an R-parity violating SUSY contribution
to the elementary vertex $\mu^- +\pi^+ \rightarrow \pi^- + \mu^+$
is presented in Fig. 2.} \label{fig.3}
\end{figure}

At the hadron level we assume dominance of the pion-exchange mode
(see Fig. \ref{fig.3}). The reasons are the same as in the case of
$0\nu\beta\beta$-decay \cite{FKSS97,FKS98a}. Enhancement of
the pion exchange mode with respect to the conventional two-nucleon
mechanism is due to the long-range character of nuclear interaction
and the details of the  bosonization of the
$\mu^- + \pi^+ \rightarrow \pi^- + \mu^+$ vertex.

The $(\mu^-,\mu^+)$ conversion rate in the considered SUSY mechanism
can be given in the form of Eq. (\ref{eq.3}) with the following LNV parameter
\begin{eqnarray}
\eta_{\lambda} = \frac{3}{8}(\eta_{T}~ + ~\frac{5}{8} \eta_{PS}).
\label{eq.20}
\end{eqnarray}
The R-parity violating parameters associated both
with gluino and neutralino exchange mechanisms are given as follows
(see \cite{FKSS97,FKS98a}):
\begin{eqnarray}
\eta_{PS} &=&  \eta_{\chi\tilde e} + \eta_{\chi\tilde f} +
\eta_{\chi} + \eta_{\tilde g} + 7 \eta_{\tilde g}^{\prime}, \\
\label{eta}
\eta_{T} &=& \eta_{\chi} - \eta_{\chi\tilde f} + \eta_{\tilde g}
- \eta_{\tilde g}^{\prime},
\label{eq.21}
\end{eqnarray}
with
\begin{eqnarray}
\eta_{\tilde g} &=& \frac{\pi \alpha_s}{6}
\frac{\lambda^{'2}_{211}}{G_F^2 m_{\tilde d_R}^4} \frac{m_p}{m_{\tilde
g}}\left[
1 + \left(\frac{m_{\tilde d_R}}{m_{\tilde u_L}}\right)^4\right],
\nonumber\\
\eta_{\chi} &=& \frac{ \pi \alpha_2}{2}
\frac{\lambda^{'2}_{211}}{G_F^2 m_{\tilde d_R}^4}
\sum_{i=1}^{4}\frac{m_p}{m_{\chi_i}}
\left[
\epsilon_{R i}^2(d) + \epsilon_{L i}^2(u)
\left(\frac{m_{\tilde d_R}}{m_{\tilde u_L}}\right)^4\right],
\nonumber \\
\eta_{\chi \tilde e} &=& 2 \pi \alpha_2
\frac{\lambda^{'2}_{211}}{G_F^2 m_{\tilde d_R}^4}
\left(\frac{m_{\tilde d_R}}{m_{\tilde e_L}}\right)^4
\sum_{i=1}^{4}\epsilon_{L i}^2(e)\frac{m_p}{m_{\chi_i}},
\nonumber \\
\eta'_{\tilde g} &=& \frac{\pi \alpha_s}{12}
\frac{\lambda^{'2}_{211}}{G_F^2 m_{\tilde d_R}^4}
\frac{m_p}{m_{\tilde g}}
\left(\frac{m_{\tilde d_R}}{m_{\tilde u_L}}\right)^2,
\nonumber \\
\eta_{\chi \tilde f} &=& \frac{\pi \alpha_2 }{2}
\frac{\lambda^{'2}_{211}}{G_F^2 m_{\tilde d_R}^4}
\left(\frac{m_{\tilde d_R}}{m_{\tilde e_L}}\right)^2
\sum_{i=1}^{4}\frac{m_p}{m_{\chi_i}}
\left[\epsilon_{R i}(d) \epsilon_{L i}(e)  + \right.
\nonumber \\
&+& \left.\epsilon_{L i}(u) \epsilon_{R i}(d)
\left(\frac{m_{\tilde e_L}}{m_{\tilde u_L}}\right)^2
+ \epsilon_{L i}(u) \epsilon_{L i}(e)
\left(\frac{m_{\tilde d_R}}{m_{\tilde u_L}}\right)^2
\right].
\label{eq.22}
\end{eqnarray}
Here, $G_F$ is the Fermi constant,
$\alpha_2 = g_2^2/(4\pi )$ and $\alpha_s = g^2_3/(4\pi )$ are
 $\rm SU(2)_L$ and $\rm SU(3)_c$ gauge coupling constants
 respectively;
$m_{{\tilde u}_L}$, $m_{{\tilde d}_R}$, $m_{\tilde g}$ and
$m_{\chi_i}$ are masses of the u-squark, d-squark, gluino and
neutralinos.  We used the neutralino couplings in the form
\cite{haberkane}:
\begin{eqnarray}
\epsilon_{L_i}(\phi ) & = & -T_3 (\phi ) {\cal N}_{i2} +
tan \theta_W [T_3 (\phi ) - Q (\phi ) ] {\cal N}_{i1},
\nonumber \\
\epsilon_{R_i}(\phi ) & = & Q (\phi ) tan \theta_W {\cal N}_{i1},
\label{eq.23}
\end{eqnarray}
where ${\cal N}_{ij}$  is  $4\times 4$ neutralino mixing matrix.

Assuming the dominance of gluino exchange, which is well motivated for
$0\nu\beta\beta$-decay \cite{FKS98a}, we obtain for the LNV parameter
in Eq. (\ref{eq.20}) the following simplified expression
\begin{equation}
\eta_{\lambda} =
\frac{\pi \alpha_s}{6}
\frac{\lambda^{'2}_{211}}{G_F^2 m_{\tilde d_R}^4}
\frac{m_p}{m_{\tilde g}}\left[
1 + \left(\frac{m_{\tilde d_R}}{m_{\tilde u_L}}\right)^2\right]^2.
\label{eta-N}
\end{equation}

We denote the $(\mu^-,\mu^+)$ conversion nuclear matrix element
to be substituted in Eq. (\ref{eq.3}) as ${\cal M}_{\lambda}$.
It can be written in the same form as for the $0\nu\beta\beta$-decay
\cite{woda,FKS98a}.
This is because for light nuclear systems the approximation used
in Eq. (\ref{eq.8}) is reasonable. Also, relatively small momentum
of the outgoing electron ($p_{\mu^+}R/3 \approx 0.1$) allows
considering only the $s_{1/2}$ electron wave state, whose radial
dependence is negligible.
Thus we have
\begin{eqnarray}
{\cal M}_{\lambda} = c_A \Big[
 \frac{4}{3}\alpha^{1\pi}\left(M_{GT}^{1\pi} + M_{T}^{1\pi} \right)
      +
      \alpha^{2\pi}\left(M_{GT}^{2\pi} + M_{T}^{2\pi} \right)\Big]\,
\label{eq.24}
\end{eqnarray}
with $c_{_{A}} = m^2_{_{A}}/(m_p m_e)$ ($m_A = 850$ MeV). The
structure coefficients of the one-pion  $\alpha^{1\pi}$ and
two-pion mode $\alpha^{2\pi}$ are \cite{FKSS97,FKS98a}:
$\alpha^{1\pi} = -0.044$ and $\alpha^{2\pi} = 0.20 $. The partial
nuclear matrix elements of the $R_p \hspace{-1em}/\;\:$  SUSY
mechanism for the $(\mu^-,\mu^+)$ process are:
\begin{eqnarray}
{M}_{GT}^{k\pi} &=&
\langle 0^+_f|~\sum_{i\neq j} ~\tau_i^+ \tau_j^+ ~
\frac{R_0}{r_{ij}}
~F_{GT}^{k\pi}(x)
~{\bf{\sigma}}_i\cdot{\bf{\sigma}}_j,
~| 0^+_i \rangle , \nonumber \\
{M}_{T}^{k\pi} &=&
\langle 0^+_f|~\sum_{i\neq j} ~\tau_i^+ \tau_j^+ ~
\frac{R_0}{r_{ij}}
~F_{GT}^{k\pi}(x)
~{\bf{S}}_{ij}
~| 0^+_i \rangle ,
\label{eq.25}
\end{eqnarray}
where $k=1,2$
and $R_0$ is the nuclear radius.
The structure functions $F_{I,J,K}$ can be expressed as:
\begin{eqnarray}
F_{GT}^{1\pi}(x_\pi ) &=& e^{-x_{{\pi}}}, ~~~~~
F_{T}^{1\pi}(x_\pi )  = (3 + 3x_{{\pi}} + x^2_{{\pi}} )
\frac{e^{-x_{{\pi}}}}{x^2_{{\pi}}}, \nonumber \\
F_{GT}^{2\pi}(x_\pi ) &=& (x_{{\pi}} - 2) ~e^{-x_{{\pi}}}, ~~~~~
F_{T}^{2\pi}(x_\pi )  = (x_{{\pi}}+1) ~e^{-x_{{\pi}}}.
\label{eq.26}
\end{eqnarray}
We used the notation:
${\bf{\hat{r}}}_{ij} =$ $(\bf{r}_i -
\bf{r}_j)/ |\bf{r}_i - \bf{r}_j|$, $r_{ij}=|\bf{r}_i - \bf{r}_j|$,
$x_{_{A}} = m_{_{A}} r_{ij}$ and $x_{\pi} = m_{{\pi}} r_{ij}$.
Here  $\bf{r}_i$ is the coordinate of the i-th nucleon.

\section{Nuclear structure calculations and numerical results}

Let us give
a short account of our approach to calculate nuclear
matrix elements of the $(\mu^-,\mu^+)$ conversion process. We used
the proton-neutron renormalized Quasiparticle Random Phase
Approximation \cite{schw,si99,woda}
and analyzed the full
$0-3\hbar\omega$ shells plus $2s_{1/2}$, $0g_{7/2}$ and $0g_{9/2}$
levels as a single-particle model space both for protons and
neutrons. The single particle energies have been obtained by using
a  Coulomb--corrected Woods--Saxon potential. Two-body G-matrix
elements we derived from the Bonn one-boson exchange potential
within the Brueckner theory. The pairing interactions have been
adjusted to fit the empirical pairing gaps \cite{cheo93}. The
particle-particle and particle-hole channels of the G-matrix
interaction of the nuclear Hamiltonian $H$ are renormalized by
introducing the parameters $g_{pp}$ and $g_{ph}$, respectively.
The calculations have been carried out for $g_{ph} = 1.0$ and
$g_{pp} = 0.8,~1.0,~1.2$. The effect of two-nucleon correlation
has been taken into account as in Refs. \cite{si99,woda}.

In Table \ref{table.1} we present the nuclear matrix elements of
the light Majorana neutrino exchange mechanism for the
$(\mu^-,\mu^+)$ conversion in $^{44}Ti$.  The adopted value of the
average nuclear excitation energy $(\bar{E}-E_i)$ is 10 MeV.  We
have found that our results depend weakly on this value within its
physical range $2~MeV~ \le (\bar{E}-E_i) \le ~15~MeV$. We used the
advantage of that the widths of the low lying nuclear states are
negligible in comparison with their energies (see comments after
Eq. (\ref{eq.7})) and carried out the calculation in the limit
$\varepsilon \rightarrow 0$. This allowed us to separate the real
and imaginary of the nuclear matrix element with the formula
\begin{equation}
\frac{1}{\alpha + i\varepsilon} =
{\cal{P}} \frac{1}{\alpha} - i \pi \delta(\alpha).
\label{eq.27}
\end{equation}

As seen from Table \ref{table.1} the $(\mu^-,\mu^+)$ conversion
nuclear matrix element significantly depends on details of the
nuclear model, in particular  on the renormalization of the
particle-particle channel of the nuclear Hamiltonian, and on the
two-nucleon short-range correlation effect (s.r.c.). Especially,
the real part of the nuclear matrix element $M^D$ exhibits strong
sensitivity to the s.r.c.. It is worthwhile to notice that the
imaginary part of the nuclear matrix element ${\cal M}_{\nu}$
dominates in $|{\cal M}_{\nu}|$, in contrast to
$0\nu\beta\beta$-decay where the imaginary part of the nuclear
matrix element is zero. The latter is a direct consequence of
special choice of nuclei for the $0\nu\beta\beta$-decay searches.
These are nuclei with energetically forbidden ordinary beta decay
channels, which, otherwise, would be the source of additional
background.

The nuclear matrix elements of heavy Majorana neutrino exchange
and $R_p \hspace{-1em}/\;\:$  SUSY mechanisms of the
$(\mu^-,\mu^+)$ conversion in $^{44}Ti$ are listed in Table
\ref{table.2}. Notice that these nuclear matrix elements are
strongly suppressed by the two-nucleon s.r.c. The trilinear
R-parity breaking matrix elements exhibit weaker dependence on the
two-nucleon s.r.c. due to the long range character of the two- and
one-pion exchange mechanisms. The sensitivity of the obtained
results to the renormalization of the particle--particle
interaction strength is not strong. According to our numerical
analysis, variation of the nuclear matrix elements presented in
Table \ref{table.2} do not exceed $20\%$ within the physical
region of the nuclear structure parameter $g_{pp}$ ($0.8~ \le
g_{pp} \le ~1.2$).

Latter on in our analysis we use the following
values of the $(\mu^-,\mu^+)$ conversion matrix elements of $^{44}Ti$
\begin{equation}
|{\cal M}_{\nu}| = 2.66, ~~~~ {\cal M}_{N} = 14.6, ~~~~
{\cal M}_{\lambda} = -426,
\label{eq.28}
\end{equation}
calculated for $g_{pp}=1.0$ taking into account the two-nucleon s.r.c.

With these values of matrix elements we
find from Eqs. (\ref{eq.3})-(\ref{eq.33}) and (\ref{eq.28})
the $(\mu^-,\mu^+)$ conversion branching ratios
\begin{eqnarray} \nonumber
&R^{(\nu)} =
1.0\times 10^{-23} \left|\frac{<m_\nu >_{\mu\mu}}{m_e}\right|^2, \ \ \ \
R^{(N)} =
3.0\times 10^{-22} \left|\eta_{N}\right|^2,&\\
&R^{(\lambda)} =
2.6\times 10^{-19}
\left|\eta_{\lambda}\right|^2.&
\label{eq.34}
\end{eqnarray}
for the above discussed light Majorana neutrino
exchange  $R^{(\nu)}$, heavy Majorana neutrino exchange
$R^{(N)}$ and the SUSY R-parity violating $R^{(\lambda)}$ mechanisms.

Eqs. (\ref{eq.34}) allow us to derive the theoretical upper bound
on the $(\mu^-,\mu^+)$ conversion branching ratio and estimate the
prospects for experimental observation of this process. Towards
this end we first find the experimental upper bounds for the LNV
parameters $\eta_i$ from other processes.

Atmospheric and solar neutrino oscillation data, combined with the
tritium beta decay endpoint, set upper bounds on the masses of the
three light neutrinos \cite{barg1} $m_{e,\mu,\tau}\leq 3$eV. Thus,
we have  conservatively
\begin{eqnarray}
\langle m_{\nu}\rangle_{\mu\mu} \leq 9~\mbox{eV}.
\label{eq.344}
\end{eqnarray}
More detailed analysis of neutrino oscillation data
in combination with the data on $0\nu\beta\beta$-decay
leads to more stringent constraints on this LNV parameter
\cite{haug}:
\begin{eqnarray}
\langle m_{\nu}\rangle_{\mu\mu} \leq 0.76~\mbox{eV}.
\label{eq. 111}
\end{eqnarray}
Assuming the existence of heavy neutrinos N, we may estimate an
upper bound on $\eta_{_N}$ using the LEP limit on heavy stable
neutral leptons $M_N\geq 39.5$ GeV \cite{LEP}, which leads to
\begin{eqnarray}
\eta{_N} \leq 2.4 \times 10^{-2}.
\label{eq.234}
\end{eqnarray}
Adopting the current upper bound $\lambda'_{211} \leq 0.059$
\cite{bar} we obtain from Eq. (\ref{eta-N}) an upper bound on the
LNV parameter $\eta_{\lambda}$ of the SUSY R-parity violating
mechanism
\begin{eqnarray}
\eta_{\lambda} \leq  6.8 \times 10^{-4},
\label{eq.356}
\end{eqnarray}
for superparticle masses $m_{\tilde g}\sim m_{\tilde q} \sim 100$
GeV.

Inserting the upper bounds on the LNV parameters
from Eqs. (\ref{eq.344})-(\ref{eq.356}) to Eqs. (\ref{eq.34}) we obtain
the following constraints
%
\begin{eqnarray}
&R^{(\nu)} ~\leq~ 2.3 \times 10^{-35} (3.2 \times 10^{-33}), \ \ \  
R^{(N)} ~ \leq ~ 1.7 \times 10^{-25},& \\ \nonumber
&R^{(\lambda)} ~ \leq~ 1.2 \times 10^{-25}.&
\label{119}
\end{eqnarray}
The value in the brackets correspond to the conservative bound
in Eq. (\ref{eq.344}). The weakest last constraint can be viewed
as an expected upper bound on the $(\mu^-,\mu^+)$ conversion branching
ratio $R \leq 1.0 \times 10^{-25}$. Note that this is about
7 orders of magnitude lower than the corresponding limit in Ref. \cite{miss}.
One of the main reasons of this difference is strong
overestimation of the corresponding nuclear matrix element in Ref.
\cite{miss} due to simplifying assumptions valid for the
$(\mu^-,e^-)$ but not for the $(\mu^-,\mu^+)$ conversion.
A particular example of such assumptions is the closure
approximation inapplicable in the latter case. In fact, studying
the $(\mu^-,\mu^+)$ conversion we deal with a two vacua problem
since the initial and final nuclear states are different.

\section{Summary}

In the present work we developed the theory for the muonic
analogue of neutrinoless double beta decay, paying special
attention to nuclear structure effects. We studied the three
presently most reliable mechanisms for $(\mu^-,\mu^+)$ conversion:
light and heavy Majorana neutrino exchange mechanisms as well as
the SUSY R-parity violating mechanism. Detailed analysis of the
corresponding nuclear matrix elements has been performed. We
uncovered the singular behavior of the nuclear matrix element of
the light Majorana neutrino exchange mechanism. These sort of
singularities are absent in the case of the
$0\nu\beta\beta$-decay. We also have shown that the imaginary part
of the matrix element of this mechanism, which was neglected
before, plays a dominant role in the numerical analysis.

We derived the $(\mu^-,\mu^+)$ conversion matrix elements for
$^{44}Ti$ within the pn-RQRPA approach,  and shown that their
values existing in the literature \cite{miss} were significantly
overestimated.

We computed the upper bounds on the branching ratio of
$(\mu^-,\mu^+)$ conversion from the current constraints on the LNV
parameters. The largest bound corresponds to the SUSY R-parity
violating mechanism $R~\leq~1.0\times 10^{-25}$ which we treat as
an expected upper bound on the branching ratio of $(\mu^-,\mu^+)$
conversion. This upper bound is about 7 order of magnitude more
stringent than those existing in the literature \cite{miss}. With
this result we conclude that the nuclear $(\mu^-,\mu^+)$
conversion is probably out of reach of the present and future
generation experiments.

\vskip10mm \centerline{\bf Acknowledgments}

This work was supported in part by the Deutsche
Forschungsgemeinschaft grant 436 SLK 17/298, by Fondecyt (Chile)
under grant 8000017 and by RFBR (Russia) under grant 00-02-17587.

\begin{table}[t]
\caption{Nuclear matrix elements of the light Majorana
neutrino exchange mechanism of the $(\mu^-,\mu^+)$ conversion in $^{44}Ti$.
The upper indices $D$ and $C$ denote the contributions associated
with the direct (Fig. 1a) and cross (Fig. 1b) Feynman diagrams,
respectively.
The calculations have been performed within pn-RQRPA without
and with taking into account the two-nucleon short-range correlations
(s.r.c.).}
\label{table.1}
\begin{tabular}{lcccccccccc}
 & \multicolumn{6}{c}{Direct terms} &
\multicolumn{3}{c}{Cross terms} &\\ \cline{2-7} \cline{8-10}

 $g_{pp}$ & $Re(M^D_F)$ & $Re(M^D_{GT})$ & $Im(M^D_F)$ & $Im(M^D_{GT})$ &
 $Re(M^D)$ & $Im(M^D)$ &
 $M^C_F$ & $M^C_{GT}$ &
 $M^C$ &  $|{\cal M}_{<m_\nu >_{\mu\mu}}|$ \\\hline
\multicolumn{11}{c}{ without s.r.c} \\
 0.80 & -0.30 &  1.23 &  1.38 & -3.10 &   1.43 & -3.99 & -0.53 &  1.33 & 1.67 &  5.05\\
 1.00 & -0.29 &  1.12 &  1.25 & -2.61 &   1.31 & -3.41 & -0.49 &  1.16 & 1.47 &  4.40\\
 1.20 & -0.28 &  1.06 &  1.12 & -2.14 &   1.24 & -2.85 & -0.44 &  1.00 & 1.28 &  3.81\\
\multicolumn{11}{c}{ with s.r.c} \\
 0.80 &  0.06 &  0.17 &  1.10 & -2.25 &   0.13 & -2.95 & -0.35 &  0.79 & 1.01 &  3.16\\
 1.00 &  0.04 &  0.17 &  0.98 & -1.83 &   0.14 & -2.46 & -0.32 &  0.67 & 0.87 &  2.66\\
 1.20 &  0.03 &  0.20 &  0.88 & -1.43 &   0.18 & -1.99 & -0.29 &  0.56 & 0.74 &  2.20\\
\end{tabular}
\end{table}

\begin{table}[t]
\caption{Nuclear matrix elements of heavy Majorana neutrino exchange
and the SUSY R-parity violating mechanisms of the
$(\mu^-,\mu^+)$ conversion in $^{44}Ti$. The shorthand s.r.c. denotes
two nucleon short range correlations.}
\label{table.2}
\begin{tabular}{lccccccccc}
 & \multicolumn{4}{c}{Heavy neutrino mech.} &
\multicolumn{5}{c}{$R_p \hspace{-1em}/\;\:$  SUSY mech.}\\ \cline{2-5} \cline{6-10}
 $g_{pp}$ & $M^{heavy}_F$ & $M^{heavy}_{GT}$ & $M^{heavy}_T$ &
${\cal M}_{\eta_{_N}^{\mu\mu}}$ & ${M}_{GT}^{1\pi}$ & ${M}_{T}^{1\pi}$ &
${M}_{GT}^{2\pi}$ & ${M}_{T}^{2\pi}$ &
${\cal M}_{{\lambda'}_{211}}^{\mu\mu}$ \\ \hline
\multicolumn{10}{c}{ without s.r.c} \\
   0.80 &  -102.  & 380. & -25.3 & 421. &  2.68 &  -1.68 &   -3.48 &  -0.696 & -1346. \\
   1.00 &  -94.8  & 344. & -24.9 & 380. &  2.39 &  -1.65 &   -3.13 &  -0.690 & -1215. \\
   1.20 &  -87.6  & 311. & -24.4 & 342. &  2.11 &  -1.61 &   -2.81 &  -0.676 &  -1094. \\
\multicolumn{10}{c}{ with s.r.c} \\
   0.80 &  -28.2 & 20.9 & -21.2 &  17.8 &  1.03 & -1.13 &  -0.930 & -0.661 &  -471. \\
   1.00 &  -26.0 & 18.9 & -21.0 &  14.6 &  0.889 &  -1.11 & -0.824 &  -0.655 & -426. \\
   1.20 &  -23.9 & 17.1 & -20.5 &  11.8 &  0.760 &  -1.09 &  -0.729 &  -0.643 & -385. \\
\end{tabular}
\end{table}


\end{document}